# Detecting Informal Organization Through Data Mining Techniques

Maryam Abdirad [1]*, Jamal Shahrabi [2]

[1] Department of Industrial, Systems and Manufacturing Engineering, College of Engineering, Wichita State University

[2] Department of Industrial Engineering, Amirkabir University of Technology, Tehran, Iran

**Abstract**

One of the main topics in human resources management is the subject of informal organizations in the organization such that recognizing and managing such informal organizations play an important role in the organizations. Some managers are trying to recognize the relations between informal organizations and being a member of them by which they could assist the formal organization development. Methods of recognizing informal organizations are complicated and occasionally even impossible. This study aims to provide a method for recognizing such organizations using data mining techniques. This study classifies indices of human resources influencing the creation of informal organizations, including individual, social, and work characteristics of an organization's employees. Then, a questionnaire was designed and distributed among employees. A database was created from obtained data. Applied data mining techniques in this study are factor analysis, clustering by K-means, classification by decision trees, and finally association rule mining by GRI algorithm. At the end, a model is presented that is applicable for recognizing the similar characteristics between people for optimal recognition of informal organizations and usage of this information.

* Corresponding author: Maryam Abdirad
mxabdirad@wichita.edu

## 1. Introduction

One of the important issues in human resource management (HRM) is the issue of formal and informal organizations in the organization. A formal organization is a conscious structure of roles in the organization that is formally organized. The description of the formal organization does not indicate that any flexible items are involved. With proper and efficient management, the formal organization is flexible.

It is noteworthy that the second aspect of formal organization in the organization is called an informal organization. The informal organization appears within the formal organization. It means that informal organization is not possible without a formal organization. An informal organization is a set of interpersonal relationships that are automatically formed in all formal organizations with the members' behavior (W. Newstrom and Davis 2006). In fact, the informal organization is formed based on shared cultural perspectives and assessments and shared interests through friendship and personal relationships. Most organizational problems for which there is no formal solution must be addressed informally. Compared with a formal organization that is planned consciously and carefully, the informal organization has a natural structure. The informal organization is usually created at the same time or after the formation of the formal organization. Therefore, the recognition and managing of informal organizations have a significant role in the management of formal organizations.

In this paper, it was decided to recognize informal organizations by data mining (DM) techniques. DM is an analytical process that is performed analysis across a large amount of data. The goal of this research is to propose a model to discover the best employee roles for each employee within the organization.

The remainder of this paper is organized as follows. Section 2 provides a theoretical foundation and background about informal organizations and DM. Section 3 explains the selected methodology in this paper. Section 4 is dedicated to a case study, including a selected company, a questionnaire for design sampling, data collection, methodology implementation, and a final model. Section 5 presents the conclusion, limits, and future directions.

## 2. Literature review

### 2.1. Informal organizations

In any formal organization, informal organizations form the most complicated system. These informal organizations are dependent on the relations between employees. For conducting tasks within the organization, both formal and informal systems are required. In this organization, managers explain the organizational relations for their employees and change such relations based on their needs (W.Newstrom 2014).

Chester Barnard is the first one who studied the relations between formal and informal organizations as well as decision processes in organization and management duties. He advocates a relationship between social and informal networks as well as with formal systems and objectives. He is one of the thinkers who discovered the importance of organizational structure and noted that informal relations help both conducting the work and assisting employees in meeting their social needs (Nikezić, Dželetović, and Vučinić 2016)(Gabor and Mahoney 2010)(Thompson Heames and Harvey 2006).

Sarlak and Salamzadeh focused on the advantages of formal and informal organizations and then analyzed the impacts of informal organizations on formal routines by measuring the communications, social capital, knowledge management, and talent management as factors of informal organizations that affect formal organizations by using deferent tests like the Kolmogorov-Smirnov and T-test. They found that there is a significant relationship between formal and informal organizations in their case study (Sarlak and Salamzadeh 2014).

Molina used social network analysis to get closer to the structure of relationships in an organization in a variety of ways like direct observation, sampling, informants, and questionnaires on relationships. This structure can resemble an informal chart (Molina 2001).

Social network analysis (SNA) is an effective and emerging tool in construction research that uses systems theory to describe how relationships influence behavior (Bonanomi et al. 2019). De Toni and Nonino determined the key roles in an informal organizational structure and outlined their contribution to the companies' performance. To find a framework, social network analysis (SNA) methodology was applied (de Toni and Nonino 2010).

Laat and Schreurs explained about their research design to find informal learning activity. As a first step, the method looks into the existence of these informal networks using social network analysis (SNA). Second, the method is used to extract what these networks are about by using group discussion software. These two steps greatly facilitate the understanding of change processes in organizations. They believed by using this approach, organizations can link with existing informal networks of practice and unlock their potential for organizational learning (De Laat and Schreurs 2013).

Morton et al. used social network analysis to find, visualize, manipulate a model for the informal organization, and develop the model into several different industrial settings. This model enables visualization of "core knowledge communities", generating discussion, and supplying focus for individuals and teams to manage relationships more effectively and hence improve product development performance (Morton et al. 2004).

Some papers applied machine learning algorithms for large networks. Puzis and Fire used classification techniques to identify informal organizations in different samples like Facebook and LinkedIn (Fire and Puzis 2016). Bonanomi et al. looked for the impact of digital technologies on organizations. This approach combined interviews, regular check-ins, and document analysis with DM and social network analysis (SNA) to find the relationships and recognize their impact on the firm's organizational structure. They created a dendrogram that shows the formal organizational structure and a sociogram that displays the informal organizational structure. They found the difference between these graphs and showed the interaction between them (Bonanomi et al. 2019).

Although managers are suspicious of informal groups and fear their potential power in penetrating between employees and controlling the organization, employees may consider it as a natural organizational phenomenon. Managers deal with employees. Becoming aware of the different roles of individuals in the groups help him to recognize that he may be admitted into one of the groups at any point and what behavior managers can expect from employees. Having such a view may help managers to accurately predict the behaviors of employees and make necessary decisions based on their behavior under different conditions (Reilly 1998).

## 2.2. Data mining

DM is a powerful science that began first in the 1990s and is the process of extracting knowledge and suitable information from databases. DM has been considered as a step in the developmental stages of knowledge extraction from known databases and has a close relationship with statistics, machine learning, database management, and model recognition, etc.(Hand, Mannila, and Smyth 2001). DM is a process applied to discover unknown relationships and data-based models (Berry and Linoff 2011). Here, some applied data mining techniques are mentioned below:

**Factor Analysis**- Factor analysis is a statistical method for analyzing the interaction between a large group of variables and describing such variables based on hidden and common aspects among existing factors. This method is mainly related to finding a way to extract the information in the main variables with the least probability of dropping information.

**Clustering** - Clustering is to divide the data into some groups in a way that different groups' data are quite different from each other and data within each group is quite similar. In clustering, the groups are not known in advance. The recognition of patterns is the main application of clustering. The applications of the clustering are in different fields such as GIS systems, image processing, medicine, bioinformatics, economic science, web browsing, etc. (Andrzej Kurgan et al. 2007).

**Classification** - Classification and the decision tree algorithm are considered as two of the methods applied to discover knowledge in the DM process. Decision trees are most able to produce perceivable rules. This makes the interpretation and interpretation of the classifications rather simple. There are different decision tree algorithms such as AID (Automatic Interaction Detection), CHAID (Chi-Squared Automatic Interaction Detection), CART (Classification and Regression Trees), OC-Tree, C4.5, C5, and QUEST(Quick Unbiased Efficient Statistical Tree) (Andrzej Kurgan et al. 2007).

**Association rule mining** (ARM) - The association algorithm is a basic DM algorithm and it is a correlation counting engine. This method discovers all possible and interesting patterns in a database. ARM can be used to discover the relationships and potential associations of attributes among huge amounts of data. These rules can be effective in finding unknown relationships and provide results that can be the basis for forecasting and decision making (Kotsiantis and Kanellopoulos 2006).

**Knowledge discovery** - The knowledge discovery process involves a set of DM techniques that aim to solve related problems, decision making based on analyzing data in a large database, finding solutions based on patterns discovered in data, and using the solution to a defined problem. A pattern reflects knowledge when:
- It is easily understandable to people
- It has high credibility
- It is highly applicable
- It can provide new knowledge for users that they did not know about or were not even looking for.

3. **Methodology**

To find a way to detect informal organizations, CRISP-DM methodology was selected. CRISP-DM is a framework for translating business problems into DM tasks. Fig. 1 shows six phases of the CRISP-DM process model including(Wirth and Wirth 2000)(Huber et al. 2019):

1. Business understanding: The start of any DM project focuses on defining project goals and changing them to a DM problem.
2. Data understanding: This step includes initial data collection, gets familiar with data, and detects interesting hypothesis from the dataset
3. Data Preparation: This step prepares data for the final dataset for the actual DM task.
4. Modeling: Models are selected and applied in this step. There are different algorithms in DM including supervised and unsupervised learning algorithms.
5. Evaluation: The built models need to be tested to ensure that they generalize against unseen data. The result determines the best models.
6. Deployment: After a successful evaluation of the trained model, it is deployed into an operating system to score or categorize new unseen data to create a mechanism for the use of that new information in the solution of the original business problem.

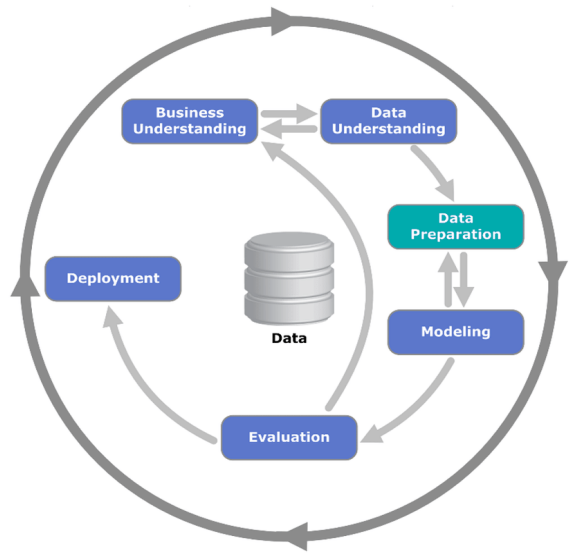

Fig. 1: Six steps according to CRISP-DM (Wirth and Wirth 2000)

## 4. Case study

According to the literature review, although using DM techniques for organizations has increased nowadays, most of these methodologies focus on limited aspects or study some cases. The proposed methodology in this research proposes combined techniques for detecting of informal organization. By using this methodology, researchers are able to present a comprehensive view of this topic.

### *4.1. SAPCO Company*

For investigating the research methodology, a case study in Supplying Automotive Parts Company (SAPCO) company in Iran has been applied. This company is one of Iran Khodro's subsidiaries, which is responsible for designing and supplying parts for Iran Khodro Company's production line. This company manages the provision of all spare parts needed, either domestic or imported, for cars produced by Iran Khodro Company. The participants include 210 employees working in the above-mentioned company.

### *4.2. Steps of the SAPCO Company case analysis*

The six steps of CRISP-DM were followed for the SAPCO company to identify informal organizations.

#### *4.2.1. Business understanding*

In the first step, the objectives of the research were defined as:

- To understand the informal organizational structure.
- To identify some similarities among different personnel.
- To select the best positions for each person in the company.

#### *4.2.2. Data understanding*

The main methods were applied to gather the required information to design a questionnaire including library-based studies, field research, and observation. The questionnaire consists of four parts (M.Huning, C.Bryan, and K.Holt 2015):

- <u>Employees General information</u>: General information are sex, age, education level, field of study
- <u>Employees Personal information</u>: Personal information are marital status, number of children, birthplace, living place, student status

- <u>Working features:</u> Some of the employee working features are working experience, year starting work, type of job, responsibility, using company transportation service, having relatives in the company, charity group membership, writing group membership, etc.
- <u>Employees' interests:</u> Some of the employee's interests are internet use, preferred film type, preferred music type, familiarity with different languages like Arabic, English, Turkish, attending sport class, attending computer class, attending music, etc.

To increase the validity of the questionnaire content, some interviews were conducted by some managers, and finally, some modifications have been made.

### 4.2.3. Data Preparation

Data preparation was performed in two stages. Based on the authors' opinion, some of the uncompleted samples were deleted. In the next stage, a test was conducted to find out whether the data was lost in the database or not. To replace the missing data, two methods named mode and average were used based on the researchers' opinion. Besides, there was no outlier since the database was provided by a person. Regarding the studies performed through Rapidminer software, the lack of outlier was best approved.

By applying factor analysis with a correlation rate of 0.4, the number of variables was reduced from 43 to 37. The deleted variables are correlated with one of the variables. The deleted similar factors will be discussed in detail as follows:

1. The working experience variable is correlated with the marital status variable and the age variable (Correlation= 0.527) and (correlation= 0.572).
2. The book-lover variable is correlated with the charity member's variable (Correlation=0.591).
3. The variable for those who are interested in a foreign football team is correlated with the variable who are interested in a national football team (correlation=0.651).
4. The variable for those who are interested in traditional music is correlated with that of those interested in all kinds of music (Correlation=0.523).
5. The variable for those who are professionally working with computers is correlated with the variable of major (Correlation=0.498).
6. The variable for those who are not participating in any class is correlated with those who are not taking part in sports classes (Correlation=0.924).

*4.2.4. Modeling and evaluation*

In this step, the actual modeling techniques are discussed. As modeling and evaluation are directly related, authors decided to combine these steps to provide a better explanation.

**Clustering**

To recognize employees with the same features and interests, the clustering technique was selected. K-Means and K-Medoids algorithms are the best known and most constructive algorithms (Yan et al. 2019). By executing the K-Means algorithm, 9 clusters were chosen as the best number of clusters. The achieved results from K-Medoids algorithms include 3 efficient clusters. The results from K-Medoids are more diverse and random. It is hard to extract clear results from it. The obtained results from the K-Means algorithm are more accurate than with K-Medoids algorithms. Also, each cluster has its special features. Through the K-Means algorithm, it can be made clear which employees belong to a given cluster. The K-Means algorithm result is briefly explained below:

- **Cluster 1:** All employees in this group work as a contractor for this company. The type of agreement for each employee causes the establishing of informal organizations, because they want to acquire information about the wages and salaries in different departments and how to get a promotion to a higher rank. Besides, these employees have no relatives in this company. This means the informal organization isn't established through family relationships.
- **Cluster 2:** The main favorites of the employees in this cluster are classic and folk music. Interest in this type of music increases conversation and the exchange of information and knowledge about music. Besides, the exchange of favorite music can establish informal groups and organizations.
- **Cluster 3:** Employees in this group are not interested in all types of music. The different opinions on special issues cause communication. Therefore, the conditions required for an informal organization will best provid for this interaction.
- **Cluster 4:** The employees in this group have a bachelor's degree. They are married. They attended the university from 2002 to 2007. They have high job qualifications and they are highly expert. They are heads or leaders of departments. They work in engineering and planning. They do not use company transportation. They are members of a charity organization. This can be considered as an opportunity for new communication to be formed centered on charity affairs. They are not interested in classical or folk music. They are not familiar with different languages.

- **Cluster 5:** The employees in this group attended the university from 2009 to 2014 for their education (bachelor's or master's degrees). They are younger than the employees in cluster 4. They spend most of their time talking about football. Their favorite team is the "Persepolis" football team (a famous football team in Iran). Such discussions on football cause the communication to be efficiently strengthened. People in this group are not members of a charity organization. All people in this group only talk in Farsi.
- **Cluster 6:** The employees in this group like sports but they do not participate in any sports classes. There are many reasons for the lack of interest in taking part in such classes, for example, lack of time or lack of friends.
- **Cluster 7:** All members of this cluster are students in a Master's degree program. To acquire information on the universities, majors, and the courses related, all students communicate with each other. They have similar positions in this company as experts. People with the same organizational status closely communicate with each other to acquire the news about the company.
- **Cluster 8:** All members in this cluster have the same education field in the industrial and engineering field. Besides, all members have the same opinions on religious affairs, which is one of the main factors in the informal organization.
- **Cluster 9:** The members in this cluster are familiar with the Arabic language. This group likes Arabic or other languages. However, this can be an option for building an informal organization.

**Classification**

The second technique applied is classification. To perform classification techniques, decision trees can be applied to the database. According to the reviewed articles, CART algorithm, CHAID algorithm, C.5 algorithm, and QUEST algorithms are considered superior to others (Lieberman and Alt 2010). The following results were achieved by comparing the trees involved:

**CART tree:** the major of education is located in the tree root. This tree includes 18 leaves such as the place of birth, duty descriptions, level of education, the entrance year to the company, and living area, located on each leaf.

**QUEST Classification Tree:** the educational major is located in the tree root. This tree has 6 leaves such as birthplace and duty descriptions are located on each leaf.

**CHAID tree:** birthplace is located in the tree root. This tree has 25 leaves such as birthplace, duty descriptions, the field of study, participation in English classes, the working year started in the company, familiarity with English language, having relatives in the company, membership in an internet group, and interest in pop music are located on each leaf.

**C5.0 tree:** birthplace is located in the tree root. It has 27 leaves on which factors such as the duty descriptions, field of study, interest in folk music, playing music instruments, and using transportation services are located on each leaf.

By comparing the obtained trees, the roots of C5.0 and CHAID trees are equal but they are different in their leaves. Also, the QUEST and Cart trees have the same roots as birthplace. The accuracy of the obtained trees is summarized in the table below.

Table 1: Comparison of the accuracy of the trees applied

| Decision tree | Accuracy |
|---|---|
| CART | 85.32% |
| C5.0 | 86.24% |
| CHAID | 85.24% |
| QUEST | 64.22% |

It appears that C5.0 algorithm has more accuracy and was chosen as the most exact one in the classification process.

**Association rule mining**

ARM is one of the important methods in the DM process (Ai et al. 2018)(Kotsiantis and Kanellopoulos 2006). ARM can extract the associations and relationships between data(Ai et al. 2018). By extraction of rules through Generalized Rule Induction (GRI) and Apriori algorithms, it was seen that the GRI algorithm provides more logical answers in the current research. The number of rules achieved through the mentioned algorithm was 100 with a confidence of 100%. Table 2 presents 8 rules for the GRI algorithm as a sample.

Table 2: Eight sample rules extracted from GRI algorithm

| Rule | Results |
|---|---|
| Their study fields are mechanical engineering, mechanical or industrial designing, civil engineering, and electronic engineering and they work as experts and they do not have any relatives in the company | They are in cluster 7. (They all are students and experts). |
| Their Birthplace is Tehran. They work as development and improvement experts. These members do not participate in any classes | They are in cluster 6. (They do not participate in any sports classes) |
| Their education levels are math and science diploma. They are familiar with the English and Arabic language | They are in cluster 9. (They are familiar with the Arabic language) |
| Their birthplace is Karaj city. They work as a contractor. They like folk music. | They are in cluster 3. (This group does not like all kinds of music) |
| Their fields of study are Executive Management, Industrial Management, Commercial Management. Their living place is the west of the city. They work as contractors. | They are in cluster 1. (All members work as contractors and they have no relatives in the company) |
| They work as experts in organizing and planning. They are fans of Iranian and European music | They are in cluster 3. (This group does not like all kinds of music) |
| Their starting work year is from 2002 to 2008 They do not use the company transportation services and are members of internet groups | They are in cluster 4. (Their entrance year to the company is from 2002 to 2007 and they are heads or leaders of departments) |
| Level of education-diploma.They are not familiar with the Arabic language and are not members of groups | They are in cluster 9. (They are not familiar with the Arabic language) |

The results of ARM were confirmed through studies and by experts in the SAPCO company. Also, the result showed that the extracted rules are directly related to the selected clusters in the clustering technique. Moreover, a comprehensive comparison of the results with the informal organization underlying assured the accuracy of the findings.

### 4.2.5. Deployment (Conceptual Model)

In this step, a model identified as the general procedure and deployment model was determined. Following the studies performed and the achieved results, the final model is presented as shown in Fig. 2.

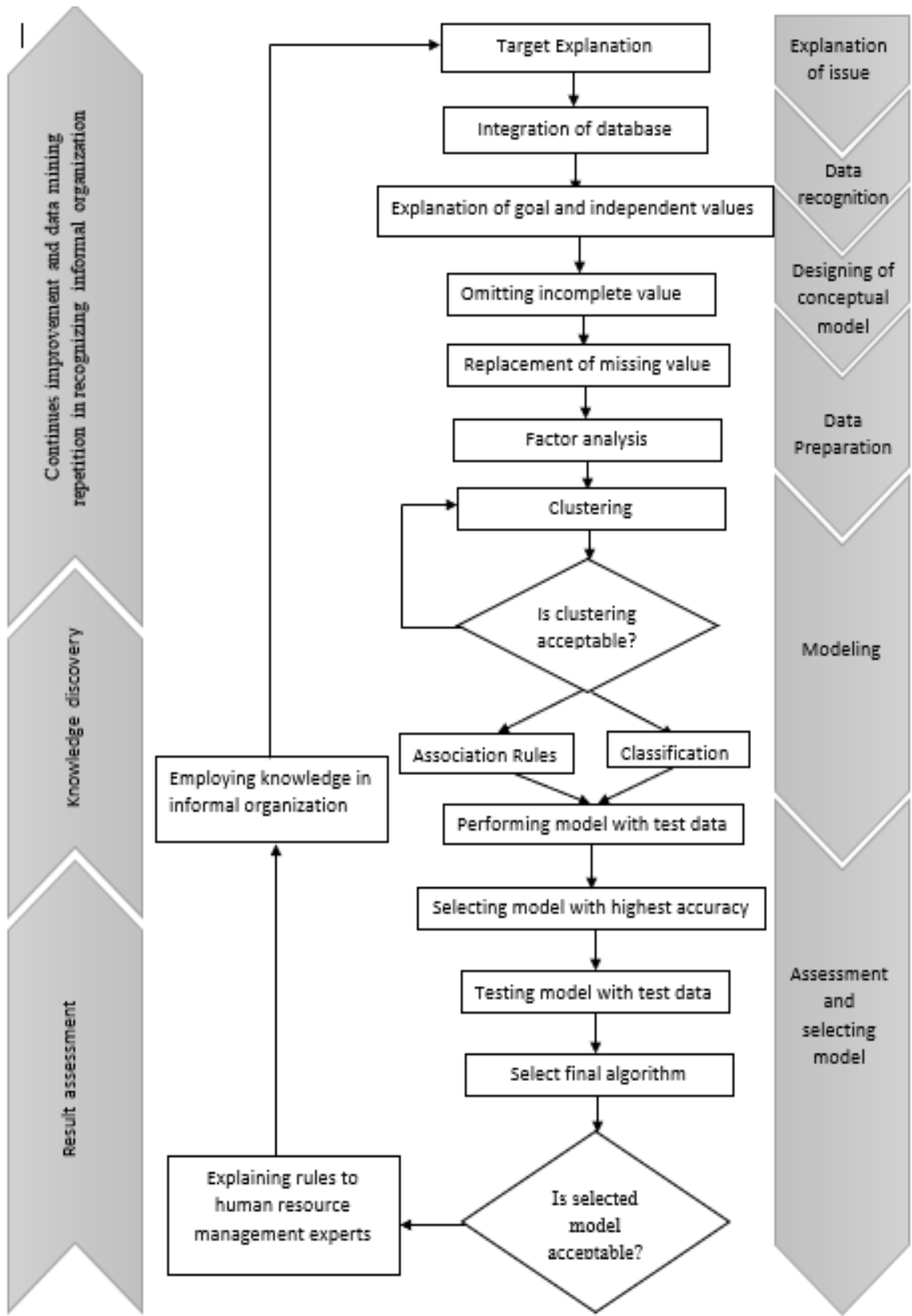

Fig. 2: The conceptual model for recognizing informal organizations through data mining

As can be seen, all the steps described in the CRISP-DM model were presented in the current model. The last step in the current model is to emphasize the importance of DM process reproducibility and to use knowledge discovery to replicate such steps. In such cases, the results can be improved. Moreover, the rules can be used every time a new employee enters the company.

## 5. Suggestion

The current research offers some suggestions for companies and organizations, managers, and future research.

It is hereby suggested to conduct research with the proposed model in the entire company with all employees' information, especially the company under the study which was previously discussed. Regarding the high association of DM methods with numbers and accuracy, developing a database can potentially promote the accuracy and exactness of the methods applied. Therefore, it is suggested to develop a model for collecting employees' information. Another suggestion is to train employees in recognizing informal organizations. It is really helpful when employees know about informal organizations as an essential requirement for each organization. Also, what the informal organizations' roles are within a company and how employees can benefit from them.

### *5.1 Managerial Implications*

Managers not only have to know about informal organizations, but they also have a fundamental need to know how to best identify and manage them. By recognizing these aspects, the managers will be able to plan and design the mutual effects and informal relationships among the employees and groups inside or outside the company. By applying the proposed model, the managers will be able to learn how to best control the positive aspects of informal organizations and potentially improve the network-based relations towards improving all functional aspects of an organization. Managers applying the proposed model, are essentially required to fully understand how, when, and where to use it. The managers are not able to recognize informal organizations by themselves (Morton et al. 2004).

Finally, the managers are expected to accept informal organizations and to attempt to perceive them as well as possible. They are essentially required to thoroughly consider their possible effects on organizations. They have to interconnects the benefits of the informal organizations with the benefits of the organization as a whole.

*5.2 Suggestions for Future Research*

More research should be done to encourage managers to build effective and constructive relationships in informal organizations, change the roles of employees in an informal organization, and develop socially constructive standards in informal organizations. In addition, it is necessary to fully consider how to control relationships and communication networks to improve performance inside and outside the organization.

This research applies to large social networks or online social networks. It is suggested to use the proposed model in large size companies, for instance Facebook and LinkedIn.

The following research fields are suggested to future researchers considering the readers and those who are most interested in the mentioned topics. The following are provided:

- Providing knowledge discovery models for recognizing informal organizations through the other clustering techniques
- Improving the model through other classification techniques like regression, support vector machines (SVM)

*5.3. Limitations*

This study has several limitations. The first limitation of this study is, employees needed to write their names on the questionnaire to detect the exact informal organization. For this reason, the number of participants was really low. Then, it was decided to remove this question from the questionnaire.

Another limitation, since some of the information in the questionnaire was personal information, was that some of the employees of the SAPCO company were not willing to cooperate in this research. Since the amount of applied data was limited, we were not able to obtain the main company network and it was not possible to fully review the informal organization network.

**6. Conclusion**

The aim of this study was to show an application of DM techniques to detect informal organizations. Considering the case study, DM can be considered as being an efficient tool for recognizing informal organizations. Factor analysis, clustering, and ARM are also considered as appropriate tools for identifying informal organizations. Therefore, the proposed model is an efficient model for informal organization recognition. By applying the mentioned techniques, organizations will be able to learn how to best control the positive aspects of informal organizations and potentially improve the network-based relations towards improving all aspects of an organization.

**References**

Ai, Dongmei et al. 2018. "Association Rule Mining Algorithms on High-Dimensional Datasets." *Artificial Life and Robotics* 23(3): 420–27. https://doi.org/10.1007/s10015-018-0437-y (August 21, 2020).

Andrzej Kurgan, Lukasz, Roman W. Swiniarski, Krzysztof Cios, and Athanasios V. Vasilakos. 2007. *Data Mining A Knowledge Discovery Approach*. ed. 1. Springer US.

Berry, Michael JA, and Gordon S Linoff. 2011. *Data Mining Techniques: For Marketing, Sales, and Customer Relationship Management*. Third. John Wiley & Sons.

Bonanomi, Marcella M. et al. 2019. "The Impact of Digital Transformation on Formal and Informal Organizational Structures of Large Architecture and Engineering Firms." *Engineering, Construction and Architectural Management* 27(4): 872–92.

Fire, Michael, and Rami Puzis. 2016. "Organization Mining Using Online Social Networks." *Networks and Spatial Economics* 16(2): 545–78. http://www.linkedin.com (August 24, 2020).

Gabor, Andrea, and Joseph T. Mahoney. 2010. Working Papers *Chester Barnard and the Systems Approach to Nurturing Organizations*. University of Illinois at Urbana-Champaign, College of Business. https://ideas.repec.org/p/ecl/illbus/10-0102.html (August 23, 2020).

Hand, David, Heikki Mannila, and Padhraic Smyth. 2001. *Principles of Data Mining*.

Huber, Steffen, Hajo Wiemer, Dorothea Schneider, and Steffen Ihlenfeldt. 2019. "DMME: Data Mining Methodology for Engineering Applications - A Holistic Extension to the CRISP-DM Model." In *Procedia CIRP*, Elsevier B.V., 403–8.

Kotsiantis, Sotiris, and Dimitris Kanellopoulos. 2006. "Association Rules Mining: A Recent Overview." *GESTS International Transactions on Computer Science and Engineering* 32: 71–82.

De Laat, Maarten, and Bieke Schreurs. 2013. "Visualizing Informal Professional Development Networks: Building a Case for Learning Analytics in the Workplace."

Lieberman, Stephen, and Jonathan Alt. 2010. "Developing Social Networks for Artificial Societies from Survey Data." In *Lecture Notes in Computer Science (Including Subseries Lecture Notes in Artificial Intelligence and Lecture Notes in Bioinformatics)*, Springer, Berlin, Heidelberg,

159–68. https://link.springer.com/chapter/10.1007/978-3-642-12079-4_21 (August 21, 2020).

M.Huning, Tobias, Phil C.Bryan, and Michael K.Holt. 2015. "Informal Social Networks in Organizations: Propositions Regarding Their Role in Organizational Behavior Outcomes." *Electronic Business Journal* 14(1).

Molina, José Luis. 2001. "The Informal Organizational Chart in Organizations: An Approach from the Social Network Analysis." *CONNECTIONS* 24(1): 78–91. http://www.sfu.ca/~insna/Connections-Web/Volume24-1/8-Molina-21.4.pdf (August 24, 2020).

Morton, S. C. et al. 2004. "Managing the Informal Organisation: Conceptual Model." *International Journal of Productivity and Performance Management* 53(3): 214–32.

Nikezić, Srđan, Milenko Dželetović, and Dragan Vučinić. 2016. "Chester Barnard: Organisational-Management Code for the 21st Century." *Procedia-Social and Behavioral Sciences* 221: 126–34. www.sciencedirect.com (August 23, 2020).

Reilly, Anthony J. 1998. *Three Approaches to Organizational Learning*. 2nd ed. eds. John E. Jones and J. William Pfeiffer. Jossey-Bass/Pfeiffer.

Sarlak, Mohammad Ali, and Yashar Salamzadeh. 2014. "Analyzing the Impacts of Informal Organizations on Formal Routing in a Networked Organization." *Journal of Asian Scientific Research* 4(12): 768–83. http://www.aessweb.com/journals/5003 (August 24, 2020).

Thompson Heames, Joyce, and Michael Harvey. 2006. "The Evolution of the Concept of the 'executive' from the 20th Century Manager to the 21st Century Global Leader." *Journal of Leadership & Organizational Studies* 13(2): 29–41. http://journals.sagepub.com/doi/10.1177/10717919070130020301 (August 23, 2020).

de Toni, Alberto F, and Fabio Nonino. 2010. "The Key Roles in the Informal Organization: A Network Analysis Perspective." *The Learning Organization* 17(1): 86–103. https://www.researchgate.net/publication/235304512 (August 23, 2020).

W. Newstrom, John, and Keith Davis. 2006. McGraw-Hill/Irwin *Organizational Behavior: Human Behavior at Work*. 12th ed. https://books.google.com/books/about/Organizational_Behavior.html?id=wZ1XAAAAYAAJ (August 24, 2020).

W.Newstrom, John. 2014. *Organizational Behavior: Human Behavior*. 14th ed. McGraw Hill.

Wirth, Rüdiger, and Rüdiger Wirth. 2000. "CRISP-DM: Towards a Standard Process Model for Data Mining." *Proceedings of the fourth international conference on the practical application of knowledge discovery and data mining*: 29–39. http://citeseerx.ist.psu.edu/viewdoc/summary?doi=10.1.1.198.5133 (August 21, 2020).

Yan, Jiali et al. 2019. "Applying Machine Learning Algorithms to Segment High-Cost Patient Populations." *Journal of General Internal Medicine* 34(2): 211–17. https://doi.org/10.1007/s11606-018-4760-8 (August 21, 2020).